\documentclass[preprint,amsmath,amssymb,superscriptaddress,apl]{revtex4}
\usepackage{rotating}
\usepackage{pstricks}
\usepackage{xcolor}
\definecolor{MyTeal}{rgb}{0 0.50.5}

\begin{document}

\title{Separation of piezoelectric grain resonance and domain wall \\dispersion
 in {Pb(Zr,Ti)O$_{\rm 3}$} ceramics}

\author{V. Porokhonskyy} \author{Li Jin
} \author{D. Damjanovic}
\affiliation{Ceramics Laboratory, Swiss Federal
  Institute of Technology (EPFL), CH-1015 Lausanne, Switzerland
}

\date{\today}

\vspace{1cm}

\begin{abstract}
  We report on the experimental investigation of a high-frequency
  (1\,MHz -- 1.8\,GHz) dielectric dispersion in unpoled and poled
  {Pb(Zr,Ti)O$_{\rm 3}$} ceramics. Two overlapping loss peaks could be
  revealed in the dielectric spectrum. The linear dependence between
  the lower-frequency peak position and average grain size
  $\overline{D}$, which holds for $\overline{D}\leqslant 10\,\mu{}m$,
  indicates that the corresponding polarization mechanism originates
  from piezoelectric resonances of grains. The intensity of the
  higher-frequency peak is drastically reduced by poling. It is thus
  proposed that this loss peak is related to domain-wall contribution
  to the dielectric dispersion.


\end{abstract}



\maketitle



It has been often observed that the lattice permittivity of
ferroelectric materials is significantly lower than the permittivity
measured at conventionally accessible frequencies, say below 1 MHz.
Perhaps the most striking manifestation of this difference is seen
in the strong, step-like dispersion of the permittivity accompanied
by the loss peak in the frequency range from 10$^8$ to
10$^{10}$\,Hz.\cite{ArltAPv3p578,McNealAPLv83p3288,KerstenFv67p191}
Above this frequency range, the permittivity is close to that of the
lattice. Several mechanisms have been proposed to explain origins of
the dispersion, including resonance of domain
walls,\cite{KittelPRv83p458} translational vibration of domain
walls,\cite{PertsevJAPv74p4105} acoustic shear waves generation by
stacks of lamellar 90$^{\circ}$-domains,\cite{ArltAPLv63p602} and
piezoelectric resonance of grains.\cite{YaoXiJACSvXp637} Some
authors have proposed that the dispersion is related to motion of
individual ions between off-center
position.\cite{MaglionePRBv40p11441} Being quite far from the state
that permits quantitative description, the proposed mechanisms are
sometimes arbitrarily invoked to explain any kind of dispersion
observed in this frequency range, thus causing
controversy and confusion. 
Most studies usually assign only one mechanism to the high
frequency dispersion in a given ferroelectric
material.\cite{ArltAPLv63p602,MaglionePRBv40p11441,KittelPRv83p458,YaoXiJACSvXp637}
However, there is no reason why several dispersive processes cannot
contribute to the permittivity simultaneously. This is particularly true
for piezoelectric resonances of grains and mechanisms related to
domain wall displacement. Domain size (distance between neighboring
domain walls) and grain size are coupled\cite{CaoJPCSV57p1499} and
their contribution to the dielectric dispersion may occur in the
narrow frequency range. The model of Pertsev and
Arlt,\cite{PertsevJAPv74p4105} for example suggests that resonance
frequency of domain walls may be lower than the frequency of the
elastic resonance of grains. The aim of this investigation is thus to
specifically address the contribution of grain resonances and domain
walls to the dielectric dispersion in 10$^8$ to 10$^{10}$\,Hz range in
technologically important ferroelectric, {Pb(Zr,Ti)O$_{\rm 3}$} or
PZT.
\par In order to obtain samples with broad range of
average grain size we take advantage of the fact that in PZT grain
size can be significantly affected by dopants.\cite{MorozovPhD}
Therefore, the samples of hard and soft PZT ceramics were prepared by
a conventional solid state process using standard mixed oxide route.
The Zr/Ti-ratio (in at.\,\%) was set to 58/42, which corresponds to a
pure rhombohedral phase at room temperature. Hardening and softening
effects were achieved by partial B-site substitution by Fe$^{3+}$ or
Nb$^{5+}$, respectively.
Further details on the sample synthesis can be found
elsewhere.\cite{MorozovPhD,MorozovJAPv104p034107}
For the sake of convenience, the samples are labeled in the following
manner: for example, PZT(58/42)Fe1.0 refers to
{Pb(Zr$_{0.58}$Ti$_{0.42}$)$_{\rm 0.99}$Fe$_{\rm 0.01}$O$_{\rm 3}$}
etc.
\par To determine grain size, the
SEM images collected on polished and thermally etched samples were
processed by means of ImageJ-software package.\cite{ImageJ}  As a
result we obtained the cross-section area on the plane of polish
$A_{cs}$ for individual grains. Further, the equivalent grain diameter
was introduced as $D_e =2\sqrt{A_{cs}/\pi}$. The distribution of $D_e$
was found to be unimodal for all compositions.  Corresponding
probability plots reveal quite good agreement with the log-normal
distribution for a number of samples, while minor right-skew
deviations were evident for some samples.  At this point one needs to
note that $A_{cs}$ and $D_e$ describe random cross-section of a grain.
It is obvious, therefore, that they underestimate its
three-dimensional extent.
To overcome this problem we adopt here the approach developed in
Ref.~\onlinecite{MendelsonJACSv52p443}. Assuming particular grain
shape, it relates average grain size $\overline{D}$ to the average
intercept length $\overline{L}$ by a proportionality constant. In the
present study 1.56 was used as such correction
factor,\cite{MendelsonJACSv52p443} which was applied to the average
value of $D_e$, instead of $\overline{L}$. It is justified in our
view, as $\overline{D}_e$ turns out to be very close to roughly
estimated $\overline{L}$, as evident from
Table\,\ref{MicroStructElMech}.  
\par For dielectric characterization,
poling and electro-mechanical measurements the samples were shaped
into the form of small cylinders, with typical dimensions of about 0.9\,mm
in diameter and 5\,mm in length. The base surfaces were electroded.
The impedance spectra in the frequency range from 1\,MHz to 1.8\,GHz
were collected by means of HP4396A Network Analyzer. 
Non-uniform fields distribution across the sample has been accounted
for during permittivity calculation.\cite{Grigas96} For each
composition the dielectric spectra were collected first starting with a
well aged unpoled state and then immediately after each poling
step.
\begin{figure}[tbp!]
\includegraphics[width=\textwidth,clip]{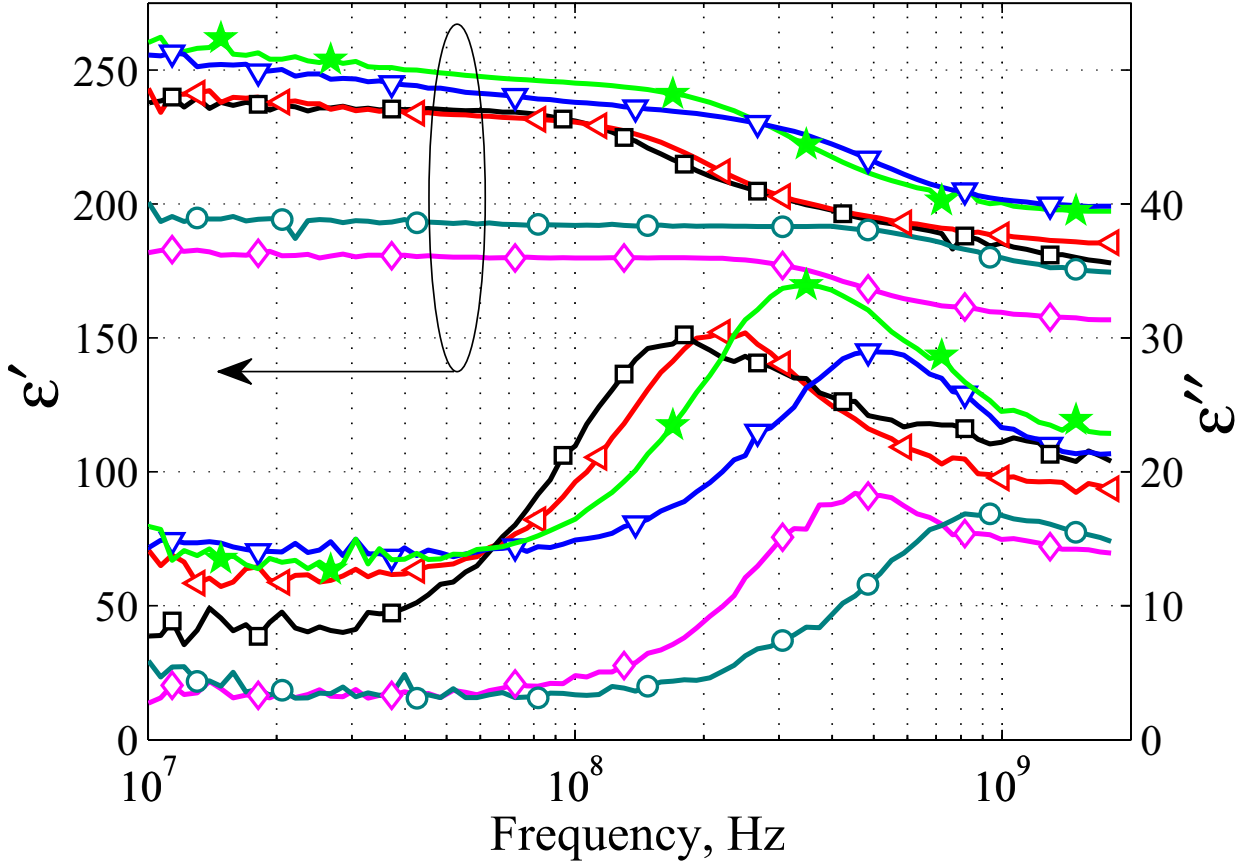} 
    \caption{\label{PoledAll} (Color online) Permittivity and loss
      spectra of well poled samples.  The symbols are introduced solely to
      facilitate distinction between different data sets: $\square$ -
      {PZT(58/42)Nb0.5}; {\red $\triangleleft$} - {PZT(58/42)Nb0.2};
      {\green $\bigstar$} - PZT(58/42)Nb0.7; {\blue $\triangledown$} -
      {PZT(58/42)Nb1.0}; {\magenta $\lozenge$} - {PZT(58/42)Fe0.5};
      \mbox{\textcolor{MyTeal}{$\circ$} - {PZT(58/42)Fe1.0}}.}
\end{figure}
\par In unpoled state all samples, soft and hard alike, exhibited
strong dispersion in agreement with earlier
studies.\cite{ArltAPv3p578,KerstenFv67p191} The loss spectra reveal
single peak located from 0.7 to about 3.4\,GHz. The upper bound of this
range was determined by means of the sleeve resonator
technique.\cite{GeyerTIMv51p383} The data are not shown here.
Hereafter the frequency of corresponding loss maximum will be referred
to as $f_{u}$.
For poling the samples were exposed to the
voltage equivalent of 25\,kV/cm at $100\div{}130^{\circ}$\,C. The
field was kept on for at least 30\,min. at elevated temperature and
during cooling.  Once the samples had been poled their
electro-mechanical properties (e.g., coupling coefficient $k_{33}$,
sound velocity $V_{s}$, etc.) were determined by means of piezoelectric
resonance method\cite{IEEEpiezoStandard} employed in the
longitudinal-bar mode. The results for $k_{33}$ (see
Tab.\,\ref{MicroStructElMech}) agree with the values typically
reported for PZT indicating that the samples were fully poled.
\begin{figure}[t!]
\includegraphics[width=\textwidth,clip]{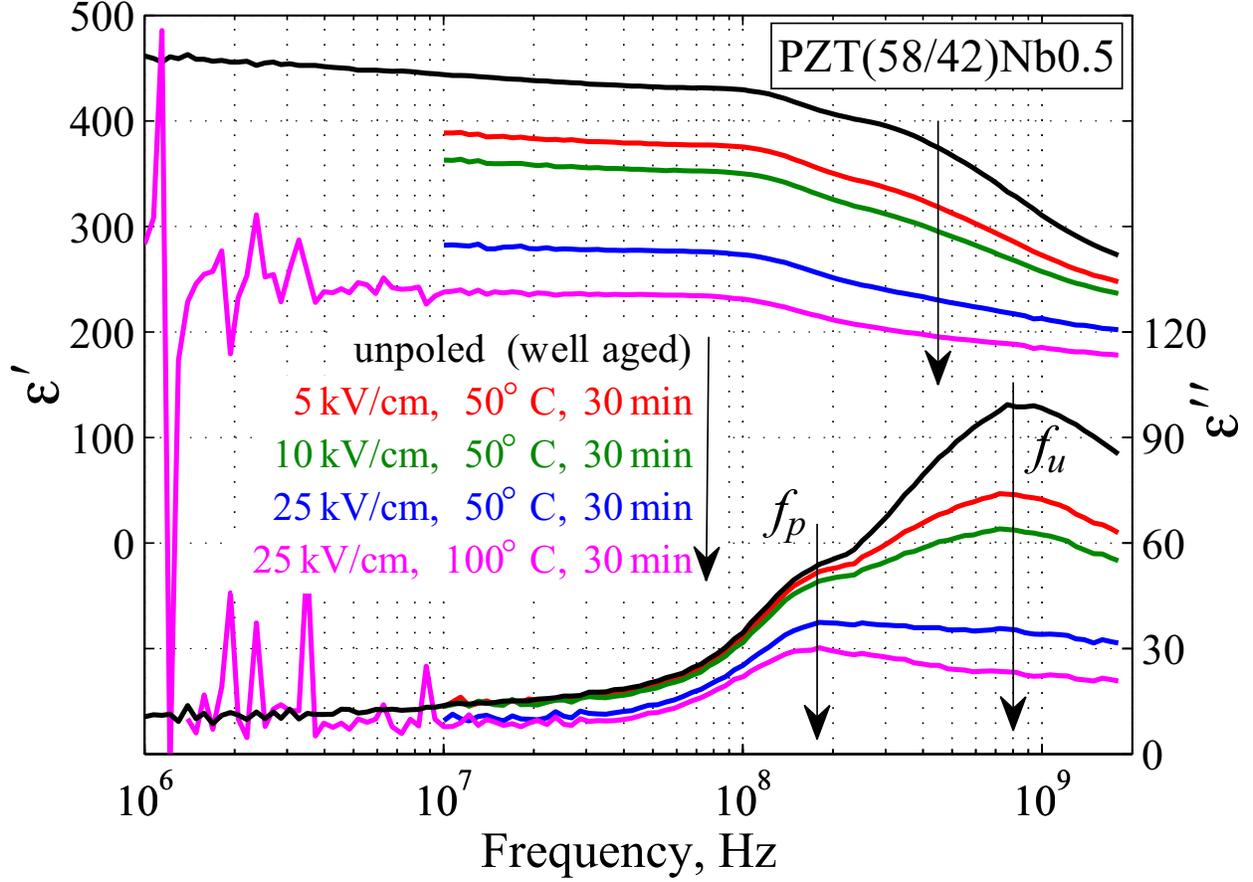} 
    \caption{\label{RN05evolutiion}(Color online) The influence of poling on
      PZT(58/42)Nb0.5 
spectra. The arrows       indicate evolution of spectra and corresponding consecutive
      poling conditions.}
\end{figure}
\par In well poled samples we observe single loss maximum. It is centered
at somewhat lower frequencies (designated here $f_{p}$) compared to
the peak position of corresponding unpoled state $f_{u}$. The
separation between $f_{u}$ and $f_{p}$ amounts to at least
0.5\,GHz. Fig.\,\ref{PoledAll} shows selected loss and permittivity
spectra of poled samples.
\begin{figure}[b!]
\includegraphics[width=\textwidth,clip]{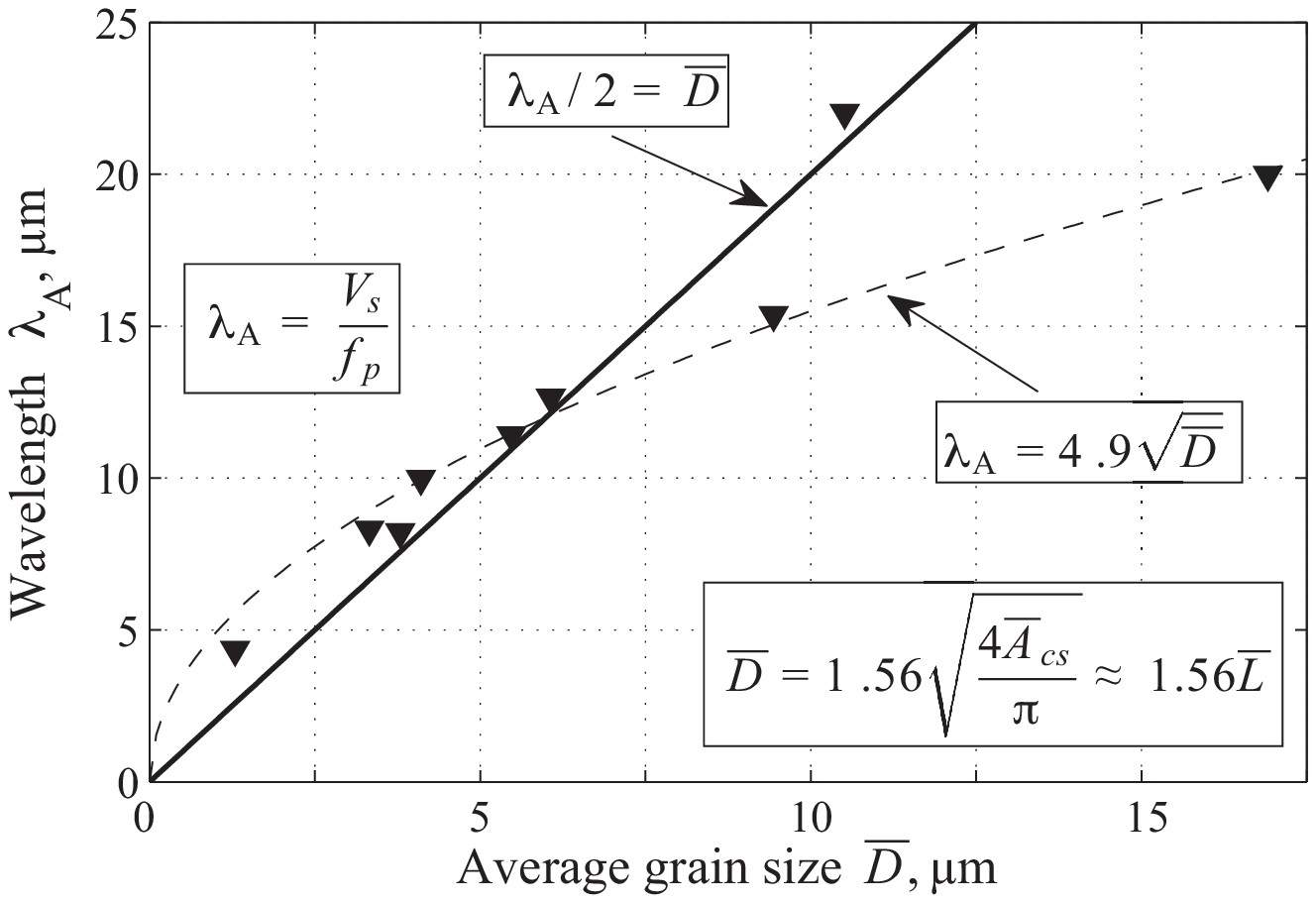} 
    \caption{\label{WavelengthSize} The acoustic wavelength $\lambda_A$, which
      corresponds to the loss peak frequency $f_{p}$, versus average
      grain size $\overline{D}$. The solid line indicates the
      calculated longitudinal resonance condition: $\overline{D}
      =\lambda_A/2$. The dashed line is the square root fit of the
      data.}
\end{figure}
To learn more about the relationship between peaks in poled and
unpoled ceramics, one of the samples, \mbox{PZT(58/42)Nb0.5}, was
poled gradually. Its dielectric spectra collected first in unpoled
state and then after each consecutive poling step are summarized in
Fig.\,\ref{RN05evolutiion}.  \begingroup \squeezetable
\begin{table*}[t!]
  \caption{\label{MicroStructElMech} Microstructural and electro-mechanical
    properties of studied PZT samples.}
  \begin{ruledtabular}
    \begin{tabular}{lcccccccccc}

      Composition & $\overline{D}_e$, $\mu{}m$ &
      $\overline{L}$, $\mu{}m$
      & $\overline{D}$, $\mu{}m$ &
      $f_{p}$, $GHz$
      & $V_s$, $m/s$ &
      $d_{33}$, $pC/N$
      & $k_{33}$ &
      $Q_m$
      & $\lambda_A$, $\mu{}m$ & Density, \%

      \\\hline
      PZT(58/42)Fe1.0 
      & 0.83 &  0.87 &
      1.3  &  0.937 &
      4073 &  77.8   &
      0.51 & 59  & 4.35  & 97.6
      \\
      PZT(58/42)Fe0.5 & 2.13 & 2.02  &
      3.32 & 0.485 &
      4028 & 57.8  &
      0.36 & 115 & 8.30 & 96.5
      \\
      PZT(58/42)Nb1.0  & 2.43 & 2.55 &
      3.79 & 0.485 &
      3987 & 102.7  &
      0.6  & 57  & 8.22  &  95.3
      \\
      PZT(58/42)Nb0.85 & 2.63 & 2.58 &
      4.1  & 0.398 &
      3971  & 100  &
      0.55  &  72  & 9.98  & 97.4
      \\
      PZT(58/42)Nb0.7 & 3.51 & 3.6  &
      5.48  & 0.348  &
      3971  & 93  &
      0.53  & 73  & 11.41   & 97.7
      \\
      PZT(58/42)Fe0.1 & 3.89 & 3.98 &
      6.21 & 0.306  &
      3871 & 80.3  &
      0.49 & 139  & 12.65 & 95.8
      \\
      PZT(58/42)       & 6.05 & 5.95 &
      9.43 & 0.267 &
      4103 & 90   &
      0.61 & 50   &  15.37    &  99.7
      \\
      PZT(58/42)Nb0.5& 6.74 & 6.71   &
      10.51& 0.180  &
      3964 & 88     &
      0.52 & 107    & 22.02    & 96.1
      \\
      PZT(58/42)Nb0.2 & 10.84 & 10.44 &
      16.91 & 0.221 &
      4415  & 77  &
      0.67  & 66  & 19.98   &  96.1
      \\
    \end{tabular}
  \end{ruledtabular}
\end{table*}
\endgroup Rapid permittivity and loss variations below 10\,MHz, which
are seen here in the last spectrum, are due to piezoelectric
resonances of the sample.  At higher frequencies each subsequent
poling step monotonously lowers permittivity and loss. More
importantly, the data indicate that the loss maximum located around
$f_{u}$ gets gradually suppressed remaining in its original position
with respect to the frequency axis. Finally, it completely disappears
from the spectrum in well poled samples. Meanwhile, the second peak
gets revealed from the feature seen as a hump on the loss spectrum of
unpoled sample. Close proximity between $f_{p}$ and the hump position
indicates that domain configuration has no influence on corresponding
polarization mechanism. The presence of such hump (seen also in
{PZT(58/42)Fe0.1}) itself suggests that two different dissipation
processes take place simultaneously in unpoled specimens.  Therefore,
we will discuss these processes separately.
 \par Let us begin with poled samples. For specimens of given composition
$f_{p}$ proved to be reproducible within the measurement
resolution and independent of sample dimensions. As $f_{p}$ tends to
be higher for the samples with finer microstructure, an obvious choice
was to look for its correlation with average grain size
$\overline{D}$.  We find it more convenient, however, to plot the
acoustic wavelength $\lambda_A$, which corresponds to $f_{p}$, as a
function of $\overline{D}$, as it accounts for the variation of sound
velocity $V_s$ 
among different compositions ($\lambda_A=V_s/f_p$). The results are
shown in Fig.\,\ref{WavelengthSize}. The solid line represents here
the longitudinal resonance condition $\overline{D}
=\lambda_A/2$.\cite{FootNote} It demonstrates reasonable agreement
with the data for $\overline{D}< 11\,\mu{}m$. It is possible to go
beyond this limit using a fit based on square root function, which is
shown in dashed line in Fig.\,\ref{WavelengthSize}. Though, it does
not provide any principal improvement over the linear dependence, as
one of data-points (at 10.5\,$\mu$m) is still found considerably
deviated from the fit. Square law could also indicate resonance of
domains, whose width is reported to be proportional to
$\sqrt{\overline{D}}$.\cite{CaoJPCSV57p1499}
We exclude
this possibility because square fit obtained here would suggest domain
width larger than $\overline{D}$.  Alternatively, one could try to
find the reason for observed deviation from the linear behavior for
{PZT(58/42)Nb0.2}, which is the composition with the largest grain
size. Indeed, there is a factor that, in our opinion, may explain it.
Namely, it is known that the grains smaller than about 10\,$\mu{}m$
tend to contain a single stack of lamellar domains, whereas larger
grains split into several
clusters.\cite{CaoJPCSV57p1499,ArltFv104p217} At present it is not
clear how well boundaries between such clusters can scatter acoustic
waves and how they are affected by poling. However, it is not
unreasonable to assume that a cluster of lamellas could behave as a
grain with effective size that is smaller than the size of the whole
grain. On condition, that this assumption is correct, $f_p$ is
expected to saturate for coarse-grain ceramics. Having that in mind,
we can draw the following conclusion. The tendency for $\lambda_A$ to
be linearly dependent on $\overline{D}$ considered together with the
fact that $f_{p}$ does not depend on the poling state (i.e. domain
pattern), indicates that the piezoelectric resonance of grains is
responsible for the loss peak observed around $f_{p}$. This conclusion
should be applied with the reservation that it concerns the ceramics
with $\overline{D}_e$ smaller than about 11\,$\mu{}m$.
\par Drastic suppression of dispersion around $f_{u}$ induced by poling
supports the idea that it is of domain-related nature. The best
explanation for such poling behavior can probably be offered by the
model of the shear wave sound emission.\cite{ArltAPLv63p602} It
essentially implies constructive interference of acoustic waves in
periodic domain structures. Poling is expected to break their
periodicity, while domain-wall density still
remains unaffected. However, other domain wall processes cannot be excluded.
\par Much higher loss intensity found around $f_{u}$ compared to the
intensity of loss peak around $f_{p}$ clearly demonstrates that the
piezoelectric resonance of grains cannot be considered as the main
mechanism responsible for strong microwave dispersion observed in
ferroelectrics. On the other hand, due to the fact that loss
associated with grain resonance dominates the loss in poled state it
can play the role of a frequency limiting factor for application
sensitive to energy loss in piezoceramics.
\par The authors acknowledge financial support of the Swiss National
Science Foundation.




\begin{thebibliography}{10}
\bibitem{FootNote} The other one-dimensional modes are
  expected to provide similar resonance criteria in given circumstances.
\end{thebibliography}

\end{document}